\documentclass[12pt]{article}
\usepackage{numbysec}
\hsize \textwidth
\advance \hsize by -\marginparwidth
\evensidemargin \oddsidemargin
\usepackage{amssymb}
\advance\hoffset by 5mm

\newtheorem{thm}{Theorem}

\newtheorem{lemma}{Lemma}

\newcommand{\pdr}[2]{\frac{\partial{#1}}{\partial{#2}}}
\newcommand{\be}{\begin{equation}}
\newcommand{\ee}{\end{equation}}
\newcommand{\la}{\label}
\newcommand{\eps}{{\varepsilon}}
\newcommand{\Erf}{{\hbox{Erf}}}
\newcommand{\commentout}[1]{}

\begin{document}
\title{Fronts in reactive convection: bounds, stability and instability}
\author{Peter Constantin\thanks{Department of Mathematics, University of 
Chicago, Chicago, IL 60637,
const@math.uchicago.edu} \and Alexander Kiselev\thanks{Department
of Mathematics, University of Wisconsin, Madison, WI 53706,
kiselev@math.wisc.edu} \and Lenya Ryzhik\thanks{Department of
Mathematics, University of Chicago, Chicago, IL 60637,
ryzhik@math.uchicago.edu}} \maketitle
\numberbysection
\begin{abstract}
We consider front propagation in a reactive Boussinesq system in
an infinite vertical strip. We establish nonlinear stability of
planar fronts for narrow domains when the Rayleigh number is
not too large. Planar fronts are shown to be linearly unstable
with respect to long wavelength perturbations if the Rayleigh
number is sufficiently large. We also prove uniform bounds on the
bulk burning rate and the Nusselt number in the KPP reaction case.
\end{abstract}

\section{Introduction}
\label{sec1} 

The presence of fluid flow may have a strong effect on reaction
processes in many situations of interest \cite{Peters,ZBLM}. Many
recent studies focused on the influence of passive advection on
combustion.
During passive advection the reacting material is carried by a
flow, but the flow is not influenced by the reaction. Traveling
waves and pulsating fronts in periodic flows have been studied in
\cite{BLL,BN,BH,X1,X2}; the thin front limit was considered in
\cite{F1,F2,MS} using homogenization techniques. Estimates on the
speed of front propagation for different classes of flows have
been derived in \cite{B1,CKOR,KR,CKR,HPS}. Numerical studies of
the effect of a passive flow on combustion were carried out in
\cite{KS,KRS,DM1,DM2,VCKRR}. This list is far from complete; we
refer to the recent reviews \cite{B1,X3} for further references.

However, there have been few rigorous works on models with
feedback of the chemical reaction on the flow field.
 In this paper we study a simplified active combustion
model \cite{MX,VR,TV1,TV2} in which the reaction does influence
the flow. The feedback of the flame on the fluid is taken in a
Boussinesq approximation. The model couples thus an
advection-diffusion-reaction equation for the temperature with an
incompressible Navier-Stokes system driven by temperature
differences. We study this problem in a two dimensional strip of
infinite vertical height and finite horizontal width. The vertical
direction is the direction of gravity. The system admits planar
fronts as particular solutions. These fronts correspond to
traveling solutions of the one dimensional reaction-diffusion
system without horizontal variation. We study them in the context
of the larger reactive Boussinesq system. Coupling with the
Boussinesq system introduces at least two new interesting effects.
One is symmetry breaking: gravity breaks the vertical symmetry of
the reaction diffusion systems. This has a dynamical effect:
fronts connecting low regions of hot fluid to high regions of cold
fluid are susceptible to the Rayleigh-Taylor instability
\cite{chandra}. The second effect is due to the introduction of
new horizontal degrees of freedom: new length scales are
introduced. When one ignores the fluid advection, the planar
fronts have a characteristic thickness $\delta$ (\ref{delta}),
which is determined by the thermal diffusivity and the
characteristic reaction time $t_c$ (\ref{tc}). Using these as
length and time units, three significant nondimensional
parameters  emerge. One is the aspect ratio $\lambda$
(\ref{lambda}), the ratio of the horizontal width of the strip to
the thickness of the planar front. The second parameter is the
Prandtl number $\sigma$ (\ref{pran}), the ratio of kinematic
viscosity to thermal diffusivity. The third important parameter is
the Rayleigh number $\rho$ (\ref{ray}) which measures the relative
strength of the buoyancy force on the scale of the front
thickness.

We prove three kinds of results: stability, instability and
uniform bounds. Stability is analyzed in Section~\ref{sec4}. If
the aspect ratio $\lambda$ is small, then the only traveling
solutions are planar fronts, and all solutions become eventually
planar. This is a consequence of the fact that diffusion acts 
rapidly across a narrow strip.
This stability mechanism is quite robust and operates for all
kinds of nonlinearities. If the Rayleigh number is large enough,
then the planar front loses stability to long wave perturbations,
which are present if the aspect ratio is large enough. The 
instability is of Rayleigh-Taylor type and is also expected to be
quite robust. We note that nonplanar traveling waves have been recently shown to exist (\cite{TV1,TV2}) in reactive Boussinesq systems with bistable
nonlinearity at large enough Rayleigh numbers. Our results on planar front
instability agree qualitatively with recent numerical results
\cite{VR}. The proof appears in Section \ref{sec5} and is based on
the intuitive idea that, in the presence of gravity, the neutral mode - corresponding to the broken vertical translation symmetry - misaligns over 
long horizontal distances, and gives rise to a long wave unstable mode. 
Technically, one needs to deal with a variable coefficient linear operator 
in which the instability of heavy fluid on top of light fluid is exploited
using the monotonicity of the front.  In Section \ref{sec3} we
prove bounds for arbitrary solutions of front type. The results
state that the bulk burning rate, speed of front, gradients of
temperature, and fluid quantities, are all bounded. The bounds do
depend on the aspect ratio. The proof applies only to concave KPP
nonlinearities. The results imply that there are no accelerating
fronts in this system. The proof of the upper bound passes via a
lower bound: the temperature gradients squared bound from below
the bulk burning rate \cite{CKOR}. On the other hand, the bulk
burning rate is bounded above by the sum of laminar burning rate
and temperature gradients to power one. This implies that the
temperature gradients are bounded, and the rest follows.

\section{Reactive Boussinesq fronts}
\label{sec2}

The reactive Boussinesq equations are
\be
\pdr{v}{t} + v\cdot\nabla v + \nabla p - \nu\Delta v = gA Te_z,
\la{ueq}
\ee
\be
\nabla\cdot v = 0,
\la{div}
\ee
\be
\pdr{T}{t} + v\cdot\nabla T -\kappa\nabla T = \frac{v_0^2}{4\kappa}f(T).\la{teq}
\ee
Here $v({\mathbf{x}},t) = (u({\mathbf{x}},t), w({\mathbf {x}},t))$ and $T({\mathbf{x}},t)$ are the velocity and the
 (normalized) temperature. The vector $e_z$ is the unit
vector pointing in the direction opposite to the pull of gravity.
The constants $\nu, \kappa, A, v_0, g$ are all positive, and
represent, respectively, the kinematic viscosity, thermal diffusivity,
thermal expansion coefficient multiplied by the temperature variation scale,
speed of purely reactive-diffusive front and acceleration of gravity.
The variables ${\mathbf{x}} = (x,z)$ belong to
 a strip, with $z\in {\mathbb R}$ and $x\in [0, L]$. The boundary conditions for the normalized temperature are
front conditions,
\be
T(x,z,t)\to 1\,\,\,\,\, {\mbox{as}}\,\, z\to -\infty, \quad T(x,z,t)\to 0 \,\,\,\,
{\mbox{as }} z\to \infty.
\la{bt}
\ee
We assume throughout the paper that the initial condition $T_0(x,z)$ satisfies
$0\le T_0(x,z)\le 1$.
The velocity vanishes at the two ends of the strip:
\be
v(x,z,t)\to 0\,\,\,\, {\mbox{as}}\,\, |z| \to\infty\la{bu}
\ee
The lateral boundary conditions are periodic:
\be
T(x + L,z,t ) = T(x,z,t),\quad v(x + L,z,t) = v(x,z,t).\la{latb}
\ee
We consider the vorticity
\be
\omega (x,z,t) = \pdr{w}{x} - \pdr{u}{z}\la{omega}
\ee
where $u$ is the component of $v$ in the $x$ direction and $w$ is the component of $v$ in
the $z$ direction. The momentum equation (\ref{ueq}) implies
\be
\pdr{\omega}{t} + v\cdot\nabla \omega - \nu\Delta\omega = gA\pdr{T}{x}\la{vort}
\ee
The nonlinearity $f(T)$ satisfies $f(0)=f(1)=0$, is a smooth function
 and is assumed to be of one of the
following three types: bistable, ignition type or concave KPP.
Bistable nonlinearities are such that there exists $\theta\in
(0,1)$ so that $f(T)<0$ for $T\in(0,\theta)$ and $f(T)>0$ for
$T\in(\theta,1)$. Ignition type nonlinearities are such that
there exists $\theta\in (0,1)$ so that $f(T)=0$ for
$T\in[0,\theta]$ and $f(T)>0$ for $T\in(\theta,1)$. Concave KPP
type nonlinearities have $f(T)>0$ for $T\in(0,1)$ and $f''(T)<-M$
for $T\in[0,1]$ with $M>0$. In their case we assume that the slope
at zero is positive and normalized, $f^{\prime}(0) = 1$, while the
slope at $1$ is negative, $f^{\prime}(1) = -m<0$. We will use the
KPP properties only in the proof of a uniform upper bound for the
burning rate presented in Section \ref{sec3}. We summarize the
essential properties of the KPP reaction used in that proof: \be
0\le f(T) \le T, \,\, -f^{\prime\prime}(T) \ge M>0, \,\,
-f^{\prime}(T)\le m,\,\,{\mbox{for}}\,\, 0\le T\le 1.\la{fineqs}
\ee

The Boussinesq system has flat traveling wave solutions
\be\la{tau} T_{fr} = \tau (z - c t),
v_{fr} = 0. 
\ee
The momentum equation (\ref{ueq}) holds in this case
because the pressure can balance  a
temperature that depends on $z$ and $t$ alone. The speed $c$ takes all values
$c\ge v_0$ in the KPP case, and is unique in the bistable and ignition case.
The profile $\tau(z)$ is monotonically decreasing in all three cases, and obeys
\[
\kappa \tau^{\prime\prime} + c\tau^{\prime} + \frac{v_0^2}{4\kappa}f(\tau) = 0,
\]
where $\tau^{\prime} = \displaystyle\frac{d{\tau}}{dz}$. We choose
now dimensional units. First we note that the function $\tau$
varies on length scale of the order \be \delta =
\frac{2\kappa}{v_0}, \la{delta} \ee that is, the function $\tau$
has the form $\tau (z) = P({z}/{\delta})$, where $P$ obeys the
equation \be P^{\prime\prime} + 2\bar cP^{\prime} + f(P) = 0.
\la{Peq} \ee Here $\bar c=c/v_0$ is a nondimensional constant
that is determined solely by the form of the nonlinearity $f(T)$,
and that in turn determines the traveling front speed $c=\bar
cv_0$ in the original dimensional variables. We choose reaction
units: we take (\ref{delta}) for length unit and we take
\be\label{tc} t_c = \frac{4\kappa}{v_0^2} \ee for time unit. The
velocity fluctuations are written as \be v({\mathbf{x}}, t) =
\frac{\delta}{t_c}{\tilde{v}}\left (\frac{{\mathbf{x}}}{\delta},
\frac{t}{t_c}\right)\la{uunits} \ee and the temperature
fluctuations are written as \be \theta(\mathbf{x}, t) =
{\tilde{\theta}}\left (\frac{{\mathbf{x}}}{\delta},
\frac{t}{t_c}\right).\la{thetaunits} \ee Using these units,
rescaling, using ${\mathbf {x}} = (x,z) = (x_{new}, z_{new}) =
(x_{old}/\delta, z_{old}/\delta)$ and $t = t_{new} =t_{old}/t_c$,
and dropping tildes, we derive the nonlinear equations \be
\pdr{\omega}{t} + v\cdot\nabla\omega -\sigma\Delta\omega =
\sigma\rho\pdr{T}{x}\la{omegane} \ee and \be \pdr{T}{t} +
v\cdot\nabla T - \Delta T = f(T)\la{teqne}, \ee where $v=(u,w)$,
with \be \Delta u=-\pdr{\omega}{z},~~\Delta w =
\pdr{\omega}{x}\la{womega} \ee which follows from (\ref{omega})
and incompressibility. The nondimensional parameters are the
Prandtl number \be \sigma = \frac{\nu}{\kappa}\la{pran} \ee and
the Rayleigh number across a laminar front width \be \rho =
\frac{gA\delta^3}{\kappa\nu}.\la{ray} \ee The boundary conditions
in $z$ are of front type
\begin{equation}
\label{bc-front}
\left \{
\begin{array}{c}
T(x,z,t)\to 1\hbox{ as $z\to -\infty$, }\\
T(x,z,t)\to 0\hbox{ as $z\to +\infty$, }\\
v(x,z,t)\to 0,\, \hbox{ as $|z|\to \infty$.}\\
\omega (x,z,t)\to 0 \hbox{ as $|z|\to \infty$.}
\end{array}
\right.
\end{equation}
The boundary conditions in $x$ are periodic
\be
T(x + \lambda, z,t) = T(x,z,t),\,\,\,v(x + \lambda, z,t ) = v(x,z,t),
\,\,\, \omega(x+\lambda,z,t) = \omega(x,z, t)\la{perniew}
\ee
with period
\be
\lambda = \frac{L}{\delta}.\la{lambda}
\ee

\section{General bounds in the KPP case}\label{sec3}

We study general solutions of the system (\ref{omegane}, \ref{teqne},
\ref{womega}) with a KPP nonlinearity that satisfies assumptions
(\ref{fineqs}), and with front-like initial data: functions that
approach the values $0$ as $z\to \infty$ and, respectively $1$ as
$z\to -\infty$ at least at an exponential rate.  We denote $D = [0,
\lambda] \times {\mathbb R}$. We will use the notation
\be
\|g\|^2_{L^2} = \frac{1}{\lambda}\int_D |g(x,z)|^2dx dz\la{ltwolam}
\ee
for the {\em normalized} $L^2$ norm. We consider the bulk burning rate
\be
V(t) = \frac{1}{\lambda}\int_D \frac{\partial T(x,z,t)}{\partial t}dx dz 
\la{v}
\ee
as a measure of the reaction rate.
Using equation (\ref{teqne}), incompressibility of $u$ and the boundary
conditions for $u$ and $T$ as $z\to\pm\infty$
we have also
\be
V(t) = \frac{1}{\lambda}\int_D f(T(x,z,t))dx dz\la{vf}
\ee
The time average of $V$ is denoted $\overline{V}$
\be
\overline{V}(t) = \frac{1}{t}\int_0^tV(s)ds.
\la{overv}
\ee

\begin{lemma} Assume that there exists a constant 
$\alpha\in{\mathbb R}$ so that
the front-like initial data $T_0(x,z)$ obeys
\be
T_0(x,z) \le \exp{(\alpha - z) }
\la{tzeroup}
\ee
and
\be
(1 - T_0(x,z)) \le \exp{(\alpha + z) }.\la{tzerodo}
\ee
Then the solution of (\ref{teqne}) obeys
the bounds
\be
T(x,z,t) \le \exp{\left [\alpha - z + 2t + 
\int_0^t\|w (\cdot, s)\|_{L^{\infty}}ds\right ]}
\la{tup}
\ee
and
\be
(1 - T(x,z,t)) \le \exp{\left[\alpha + z + t -
 \int_0^t\|w (\cdot, s)\|_{L^{\infty}}ds\right ]}\la{tdo}
\ee
for all $t\ge 0$.
\end{lemma}

\noindent{\bf Proof.}
For the bound of $T$ we seek a supersolution of the form:
$$
\theta_{+}(z,t) = \exp{[-az + 
a\int_0^{t}\|w(\cdot,s)\|_{L^{\infty}}ds + bt + \alpha]},
~~a>0.
$$
Using (\ref{fineqs}) we get
$$
\pdr{\theta_{+}}{t} + v\cdot\nabla {\theta_{+}} -\Delta\theta_{+} - f(\theta_{+})\ge 0
$$
if $b\ge a^2+1$. We chose for simplicity of exposition $a=1$, and took
the most economic $b=2$.
For the bound of $1-T$ we seek a subsolution for $T$ of the form
$$
\theta_{-}(z,t) = 1-\exp{[az-a\int_0^t\|w(\cdot, s)\|_{L^{\infty}}ds
+ bt+\alpha]}
$$
and, using the fact that $f\ge 0$ on $[0, 1]$, the condition
$$
\pdr{\theta_{-}}{t} + v\cdot\nabla {\theta_{-}} -\Delta\theta_{-}
-f(\theta_{-})\le 0
$$
follows if
$$
b\ge a^2.
$$
Again, we chose $a=1$ for simplicity, and took the best $b=1$.
The proof shows that similar inequalities  hold if
the exponential rate of decay at infinity of the initial data is
different than $a=1$, or if the initial data approaches its limit at
negative infinity at a different exponential rate than $a=1$.

Let us consider the average quantities
\be
W(t) = \frac{1}{t}\int_0^t\|w(\cdot,s)\|_{L^{\infty}}ds\la{wt}
\ee
and
\be
N(t) = \frac{1}{t}\int_0^t\|\nabla T (\cdot, s)\|_{L^2}^2ds\la{nt}
\ee
representing the average maximum vertical velocity and temperature gradient
squared.
\begin{lemma} Consider front-like initial data
that satisfy (\ref{tzeroup}). Then the solutions of (\ref{teqne})
obey
\be
\overline{V}(t) \le W(t) + 2 + \frac{\gamma}{t}\la{vineq}
\ee
for all $t\ge 0$ with $\gamma$ depending on the initial data.
\end{lemma}

\noindent{\bf Proof.}
First we write
\[
\overline{V}(t) = \frac{1}{\lambda t}\int_D\left (T(x,z,t) - T_0(x,z)\right)dx dz .
\]
which we bound as
\be
\overline{V}(t) \le  \frac{1}{\lambda t}\int_0^{\lambda} dx 
\left [\int_{-\infty}^{0}
\left (1 - T_0(x,z)\right)dz + \int_0^{\infty}T(x,z,t)dz\right ],\la{vbri}
\ee
using the fact that $T(t,x,z)\le 1$.
Now, denoting
\be
B_1(t) = \alpha + 2t+\int_0^t\|w(\cdot, s)\|_{L^{\infty}}ds\la{At},
\ee
we have from (\ref{tup}) that
$$
\int_{B_1(t)}^{\infty}T(x,z,t)dz \le 1,
$$
while, because $T\le 1$, we have
$$
\int_0^{B_1(t)} T(x,z,t)dz \le B_1(t).
$$
We use these bounds in (\ref{vbri}) and obtain (\ref{vineq}).
Now we bound $W(t)$ below by $N(t)$:
\begin{lemma} Consider  front-like initial data
that satisfy (\ref{tzeroup}). Then the solutions of (\ref{teqne})  obey
\be
N(t) \le \frac{m+2}{M}W(t) + \frac{2m+3}{M} + \frac{\Gamma}{t}\la{nw}
\ee
with $m$ and
$M$  given in (\ref{fineqs}) and with $\Gamma$ depending on the initial data.
\end{lemma}

\noindent{\bf Proof.} We start by computing
$$
\frac{d}{dt}V(t) = \frac{1}{\lambda}\int_D f^{\prime}(T(x,z,t))\frac{\partial T(x,z,t)}{\partial t}dx dz.
$$
Using (\ref{teqne}) and integrating by parts, using incompressibility
of $u$ and the boundary conditions, we obtain
\[
\frac{dV}{dt} - \frac{1}{\lambda}\int_Df^{\prime}(T(x,z,t))f(T(x,z,t))dx dz =
-\frac{1}{\lambda}\int_Df^{\prime\prime}(T(x,z,t))|\nabla T(x,z,t) |^2dx dz.
\]
Using (\ref{fineqs}) we deduce
$$
\frac{dV}{dt} + m V(t) \ge M \|\nabla T(\cdot, t)\|^2_{L^2}.
$$
Taking a time average we get
\begin{equation}
\frac{1}{t}\left (V(t) - V(0)\right) + m\overline{V}(t) \ge M N(t).
\la{vnint}
\end{equation}
We observe that
\begin{eqnarray}\label{vthree}
&&V(t) = \frac{1}{\lambda}\int_D f(T(x,z,t))dx dz =
 \int_0^{\lambda}\frac{dx}{\lambda}\int_{-\infty}^{-B_2(t)}f(T(x,z,t))dz
\\ &&~~~~~~~
+  \int_0^{\lambda}\frac{dx}{\lambda}\int_{-B_2(t)}^{B_1(t)}f(T(x,z,t))dz
+  \int_0^{\lambda}\frac{dx}{\lambda}\int_{B_1(t)}^{\infty}f(T(x,z,t))dz,\nonumber
\end{eqnarray}
where $B_1(t)$ is given in (\ref{At}) and
\[
B_2(t) = \alpha + t + \int_0^{t}\|w(\cdot, s)\|_{L^{\infty}}ds.
\]
We use the inequality
$f(T) \le m(1-T)$
which follows from (\ref{fineqs}). Then we have
$$
\int_0^{\lambda}\frac{dx}{\lambda}\int_{-\infty}^{-B_2(t)}f(T(x,z,t))dz \le m,
$$
as follows from (\ref{tdo}). Similarly,
$$
\int_0^{\lambda}\frac{dx}{\lambda}\int_{B_1(t)}^{\infty}f(T(x,z,t))dz \le 1
$$
follows from (\ref{tup}) and (\ref{fineqs}). The second term on the right in
(\ref{vthree}) is bounded
by $B_1(t) + B_2(t)$. Thus, returning to (\ref{vnint}),
we have
\[
MN(t) \le m\overline{V}(t) + 3 + 2W(t) + \frac{c}{t}.
\]
In view of (\ref{vineq}) of the previous lemma, (\ref{nw}) is proven.

The next step consists in bounding the quantity $W(t)$ in terms of $N(t)$, using the vorticity equation
(\ref{omegane}).
\begin{lemma}\label{lemma4} There exists an absolute constant
$C$ so that, for all $t>0$ one has
\be
W(t) \le C{\lambda}^{3/2}\left\{\rho\lambda{\sqrt{N(t)}} + \frac{1}{\sqrt{\sigma t}}\|\omega_0\|_{L^2}\right\}\la{wi},
\ee
where $\omega_0(x,z)$ is the initial data for $\omega(x,z,t)$.
\end{lemma}

\noindent{\bf Proof.} We multiply (\ref{omegane}) by $\omega$ and integrate.
After one integration by parts
we obtain
\begin{equation}
\frac 12\frac{d}{dt}\int_D|\omega(x,z,t)|^2\frac{dxdz}{\lambda}  +
\sigma \int_D|\nabla\omega(x,z,t)|^2\frac{dxdz}{\lambda}  =
\sigma\rho\int_D\omega(x,z,t)\pdr{T(x,z,t)}{x}\frac{dxdz}{\lambda}.
\la{balone}
\end{equation}
We introduce
\[
{\overline{T}}(z,t) : = \int_0^{\lambda}T(x,z,t)\frac{dx}{\lambda}
\]
and note the obvious fact that
$$
\pdr{T(x,z,t)}{x} = \pdr{\left (T(x,z,t) - {\overline{T}}(z,t)\right)}{x}.
$$
Inserting this expression into
the right hand side of (\ref{balone}) and integrating by parts on the right
side we obtain
\begin{eqnarray*}
&&\frac 12\frac{d}{dt}\int\limits_D|\omega(x,z,t)|^2\frac{dxdz}{\lambda}  +
\sigma \int\limits_D|\nabla\omega(x,z,t)|^2\frac{dxdz}{\lambda} \\
&&=
-\sigma\rho\int\limits_D\pdr{\omega(x,z,t)}{x}
\left(T(x,z,t) - {\overline{T}}(z,t)\right)
\frac{dxdz}{\lambda}
\end{eqnarray*}
Using Young's inequality together with the inequality
\[
\int_D\left | T(x,z,t) - {\overline{T}}(z,t)\right |^2\frac{dxdz}{\lambda}
\le \lambda^2\int_D\left | \nabla T(x,z,t) \right |^2\frac{dxdz}{\lambda}
\]
we deduce
\begin{eqnarray}
&&\frac12\frac{d}{dt}\int_D|\omega(x,z,t)|^2\frac{dxdz}{\lambda}  +
\sigma \int_D|\nabla\omega(x,z,t)|^2\frac{dxdz}{\lambda}
\nonumber\\
&&\le
\frac{\sigma}{2}\int_D\left |\pdr{\omega(x,z,t)}{x}\right |^2
\frac{dxdz}{\lambda}
+ \frac{\sigma\rho^2\lambda^2}{2}\int_D
\left | \nabla T(x,z,t) \right |^2\frac{dxdz}{\lambda}.
\la{balthree}
\end{eqnarray}
Integrating (\ref{balthree}) in time we deduce
\be
\frac{1}{t}\int_0^t ds\int_D\left |\nabla \omega (x,z,s)\right |^2
\frac{dxdz}{\lambda}
\le \rho^2\lambda^2 N(t) + \frac{1}{\sigma t}\|\omega_0\|_{L^2}^2\la{gineq}
\ee
Let us represent the function $w$ in terms of its Fourier series
\[
w(x,z,t) = \sum_{k\in \frac{2\pi}{\lambda}{\mathbf Z}}w_k(z,t)e^{i kx}
\]
and note that, in view of incompressibility, $w_0(z,t)$ is independent of $z$,
and hence the boundary conditions at $z\pm\infty$ imply that
$$
w_0(z,t) = 0.
$$
In view of (\ref{womega}), the embedding inequality
\be
\|w (\cdot, t)\|_{L^{\infty} } \le 
C\lambda^{3/2}\|\nabla\omega (\cdot,t)\|_{L^2 }\la{emb}
\ee
follows. The constant $C$ is absolute (recall that the $L^2$ norm is normalized
by $\lambda$ (\ref{ltwolam})); it can be computed either by using
Parseval's identity or using (\ref{el}) below. From (\ref{emb}) and 
(\ref{gineq}) we deduce
(\ref{wi}), using the Cauchy-Schwartz inequality for integration in time.

\begin{thm} Solutions of (\ref{omegane}), (\ref{teqne}), (\ref{womega}) 
with front-like initial data obey
\be
N(t) \le \frac{C^2(m+2)^2}{M^2}\rho^2\lambda^5 + \frac{4m+6}{M} +
(\frac{K_1}{{\sqrt{t}}} + \frac{K_2}{t}),\la{ni}
\ee

\be
W(t) \le \frac{C^2(m+2)}{M}\rho^2\lambda^5 +
C\sqrt{\frac{4m+6}{M}}\rho\lambda^{5/2} + \frac{K_3}{t^{\frac{1}{4}}}
+ \frac{K_4}{t^{\frac{1}{2}}}\la{winq}
\ee
and
\be
\lim\sup_{t\to\infty}{\overline{V}}(t) \le 2 + \frac{C^2(m+2)}{M}\rho^2
\lambda^5 + C{\sqrt{\frac{4m + 6}{M}}}\rho\lambda^{5/2}
\la{vend}
\ee
with $M, m$ given in (\ref{fineqs}), $C$ the absolute constant of
(\ref{emb}) and with $K_1, K_2, K_3, K_4$ depending on the initial
data .
\end{thm}

\noindent{\bf Proof.} We insert (\ref{wi}) into the  right hand side of
(\ref{nw}). We obtain $$ N(t) \le
\frac{m+2}{M}C\rho\lambda^{5/2}{\sqrt{N(t)}} +
\frac{C\lambda^{3/2} (m+2)}{M{\sqrt{\sigma t}}}\|\omega_0\|_{L^2}+ 
\frac{2m+3}{M} + \frac{\Gamma}{t}.
$$
Now we use Young's inequality
$$
\frac{m+2}{M}C\rho\lambda^{5/2}{\sqrt{N(t)}}\le \frac{N(t)}{2} +
\frac{C^2(m+2)^2}{2M^2}\rho^2\lambda^5
$$
and deduce (\ref{ni}). Then (\ref{winq}) follows from (\ref{ni})
and (\ref{wi}), while (\ref{vend}) follows from (\ref{winq}) and
(\ref{vineq}).

\noindent {\bf Remarks.} 1. In terms of the original dimensional
variables $x_{old}$ of (\ref{ueq}),(\ref{div}),(\ref{teq}) the
bound on the wave number (\ref{ni}) means that the Nusselt number
on scale $L$
$$
Nu = \lim\sup_{t\to\infty}\frac{1}{t}\int_0^t\int_0^L\int_{-\infty}^{\infty}
|\nabla T(x,z,s)|^2dxdzds = \lambda\lim\sup_{t\to\infty}N(t)
$$
is bounded as \be Nu \le \frac{C^2(m+2)^2}{M^2}Ra^2 +
\frac{4m+6}{M}\lambda \la{no} \ee with $Ra$ the Rayleigh number on
scale $L$: \be Ra = \frac{gAL^3}{\nu\kappa}.\la{ra} \ee

\noindent 2. Numerical evidence \cite{VR} shows that the
$\overline{V}(t)$ is indeed an increasing function of $\lambda$
and $\rho.$

\section{Nonlinear stability of planar fronts in narrow
domains}\label{sec4}


In this section we consider the reaction-diffusion Boussinesq
problem (\ref{omegane}) - (\ref{womega}) in a narrow domain, i.e.
for small aspect ratio $\lambda$. The nonlinearity $f$ is of
either one of the three types: KPP, ignition or bistable. We keep
the time and space units chosen above (\ref{delta}, \ref{tc},
\ref{uunits}, \ref{thetaunits}). We prove two results. The first
one concerns traveling solutions of the form
\begin{equation}\label{tr-front}
T(x,z,t)=T(x,z-ct),~~v(x,z,t)=v(x,z-ct).
\end{equation}
The result states that, if the Rayleigh number $\rho$ is sufficiently small,
then such solutions must be planar fronts. Planar fronts are solutions of the
form
\be
T(x,z,t) = P(z-ct), ~~ v(x,z,t) = 0\la{tauu}
\ee
for which $T$ does not depend on $x$ and $v$ vanishes.
Planar fronts do exist for $c\ge 2$ in the KPP
nonlinearity case and for a unique front speed $c_*$ in the ignition and bistable cases.

\begin{thm}\label{thm2}
There exist constants $C_1>0$ and $C_2>0$ such that if
$\lambda<C_1$, and $\rho<C_2/\lambda^3$, then the only solutions
of (\ref{omegane})-(\ref{womega}), (\ref{bc-front}, \ref{perniew})
of traveling front type (\ref{tr-front}), are planar fronts of
the form (\ref{tauu}).
\end{thm}
The second result in this section is about arbitrary solutions. We show that
all solutions of the Boussinesq system in a narrow domain become eventually
planar:
\begin{thm}\label{thm3}
There exist constants $C_1>0$ and $C_2>0$ so that if $\lambda<C_1$ and
$\rho<C_2/\lambda^3$, then
\begin{equation}\label{eq-thm3}
\|\omega(\cdot, t)\|_{L^2}+\|T_x(\cdot, t)\|_{L^2}\to 0~\hbox{ as $t\to +\infty$.}
\end{equation}
Moreover, the front speed is uniformly bounded:
\begin{equation}\label{kpp-upperbd}
\limsup_{t\to +\infty}\bar V(t)\le 2.
\end{equation}
\end{thm}
{\bf Proof of Theorem \ref{thm2}.} Let $T(x,z-ct)$, $v(x,z-ct)$ be
a traveling front solution of (\ref{omegane})-(\ref{womega}),
then
\begin{eqnarray}\label{tr-front-eqs}
&&-cT_z+v\cdot\nabla T=\Delta T+f(T)\\
&&-cv_z+v\cdot\nabla v+\nabla p=\sigma\Delta v+\sigma\rho T\hat e_z\nonumber\\
&&-c\omega_z+v\cdot\nabla\omega=\sigma\Delta\omega+\sigma\rho T_x.\nonumber
\end{eqnarray}
We multiply the second equation in (\ref{tr-front-eqs}) by $v$ and integrate
to obtain
\[
\sigma\|\nabla v\|^2=\sigma\rho\int T(x,z)w(x,z)dxdz.
\]
In this section we will use the plain $L^2$ norm
$$
\|g\|^2 = \int_D |g(x,z)|^2dxdz.
$$
The right side may be bounded as in the proof of Lemma \ref{lemma4}  to obtain
\begin{equation}\label{2.1}
\sigma\|\nabla v\|^2=\sigma\rho\int T(x,z)w(x,z)dxdz
\le \sigma\rho\|T-\overline T\| \|w\|\le
\sigma\rho\lambda\|T_x\| \|w\|,
\end{equation}
where
\[
\overline T(z)=\int_0^\lambda T(x,z,t)\frac{dx}{\lambda}.
\]
We used above the fact that $w(x,z)$ has mean zero in $x$ because of incompressibility
and boundary conditions at $z\to\infty$. Furthermore,
it satisfies the Poincar\'e inequality so that $\|w\| \le C\lambda\|\nabla
v\|$, and the incompressibility of $v$ implies that $\|\nabla
v\| =\|\omega\|$.  Therefore (\ref{2.1}) implies that
\[
\|w\| \le {C\rho\lambda^3}\|T_x\|
\]
and
\begin{equation}\label{2.2}
\|\omega\| \le {C\rho\lambda^2}\|T_x\|.
\end{equation}
We differentiate the first equation in (\ref{tr-front-eqs}) in $x$,
multiply by $T_x$ and integrate to obtain:
\begin{equation}\label{2.3}
-\int T_x(v_x\cdot\nabla T)dxdz=\int|\nabla T_x|^2dxdz-
\int f'(T)T_x^2dxdz.
\end{equation}
We use the Poincar\'e inequality for $T_x$ to obtain a bound for the
right side:
\begin{eqnarray*}
&& \int|\nabla T_x|^2dxdz-
\int f'(T)T_x^2dxdz\ge
\left(1-{C}{\lambda^2}\right)\int|\nabla T_x|^2dxdz\ge
\frac{1}{2}\|\nabla T_x\|^2
\end{eqnarray*}
if $\lambda\le C_1$ is sufficiently small, with $C_1$ an absolute constant
that may depend only on $m_1=\max_{0\le T\le 1}|f'(T)|$.
The left side of (\ref{2.3})
is bounded by
\[
\left|\int T_x(v_x\cdot\nabla T)dxdz\right|=
\left|\int T(v_x\cdot\nabla T_x)dxdz\right|\le \|v_x\|\|\nabla T_x\|\le
\|\omega\|\|\nabla T_x\|.
\]
The last two bounds imply that $\|\nabla T_x\|\le 2\|\omega\|$ for
a sufficiently small $\lambda$. Then the inequality (\ref{2.2}) and the Poincar\'e
inequality for $T_x$ imply that
\[
\|\nabla T_x\| \le{C\rho \lambda^2}\|T_x\| \le
{C\rho \lambda^3}\|\nabla T_x\|.
\]
Therefore there exists a constant $C_0>0$ such that no
$x$-dependent traveling front exists if
\begin{equation}\label{R-bd}
\rho\le \frac{C_0}{\lambda^3}.
\end{equation}
This finishes the proof of Theorem \ref{thm2}. We now
prove Theorem \ref{thm3}.

{\bf Proof of Theorem \ref{thm3}.}
We multiply the evolution equation for $v(x,z,t)$ by $v=(u,w)$ and integrate
to obtain
\begin{equation}\label{u-l2}
\frac{1}{2}\frac{d}{dt}\left\|v(\cdot, t)\right\|^2+\sigma\|\nabla v(\cdot, t)\|^2 =
\sigma\rho\int T(x,z,t)w(x,z,t)dxdz.
\end{equation}
The right side of the above equation may be bounded as in (\ref{2.1}).
We integrate (\ref{u-l2}) in time, and use (\ref{2.1}) and
the Poincar\'e inequality for $w(t,x,z)$ to get
\begin{eqnarray}
&&\left\|w(\cdot, t)\right\|^2+\frac{C\sigma}{\lambda^2}\int_0^t\|w(\cdot, s)\|^2ds\le
\sigma\rho\lambda\int_0^t\|T_x(\cdot, s)\|\|w(\cdot, s)\| ds+\left\|v(\cdot, 0)\right\|^2\\
&&\le  \sigma\rho\lambda\left(\int_0^t\|T_x(\cdot, s)\|^2ds\right)^{1/2}
\left(\int_0^t\|w(\cdot, s)\|^2ds\right)^{1/2}+\left\|v(\cdot, 0)\right\|^2\nonumber\\
&&\le \frac{\sigma\rho\lambda}{2}\left(\frac{C}{\rho\lambda^3}\int_0^t\|w(\cdot, s)\|^2ds+
\frac{\rho\lambda^3}{C}\int_0^t\|T_x(\cdot, s)\|^2ds\right)+
\left\|v(\cdot, 0)\right\|^2\nonumber
\end{eqnarray}
Therefore we have
\begin{eqnarray*}
&&\int_0^t\|w(\cdot, s)\|^2ds\le
{C\rho^2\lambda^6}\int_0^t\|T_x(\cdot, s)\|^2ds+\frac{C\lambda^2}{\sigma}
\left\|v(\cdot, 0)\right\|^2.
\end{eqnarray*}
Furthermore, once again, after integrating (\ref{u-l2}) in time we obtain
\begin{eqnarray}\label{bd-omegal2}
\int_0^t\|\omega(\cdot, s)\|^2ds\le {\rho\lambda}\int_0^t\| T_x(\cdot, s)\|
\|w(s)\|ds\le{C\rho^2\lambda^4}\int_0^t\|T_x(s)\|^2ds+
\frac{C}{\sigma}\|v(\cdot, 0)\|^2.
\end{eqnarray}
Furthermore, we differentiate the first equation in (\ref{teqne}) in $x$
and obtain
\begin{equation}\label{2.5}
T_{xt}+v\cdot\nabla T_x+v_x\cdot\nabla T=\Delta T_x+f'(T)T_x.
\end{equation}
Then, as in the proof of Theorem \ref{thm2}, provided that $\lambda<C_1$
is sufficiently small, if we multiply (\ref{2.5}) by $T_x$, and integrate,
we obtain, using the Poincar\'e inequality for $T_x$:
\[
\frac 12\frac{d}{dt}\left(\|T_x\|^2\right)+C\|\nabla T_x\|^2\le
\|v_x\|\|\nabla T_x\|.
\]
This implies that
\[
\frac 12\frac{d}{dt}\left(\|T_x\|^2\right)+\frac{C'}{2\lambda^2}\|T_x\|^2\le
\frac{C''}{2}\|\omega\|^2
\]
and hence
\begin{equation}\label{2.6}
\|T_x(\cdot, t)\|^2\le \|T_x(\cdot, 0)\|^2e^{-C't/\lambda^2}+
C''\int_0^t\|\omega(\cdot, s)\|^2e^{-C'(t-s)/\lambda^2}ds.
\end{equation}
We integrate (\ref{2.6}) and obtain
\begin{eqnarray*}
&&\int_0^t\|T_x(\cdot, s)\|^2ds\le  {C\lambda^2}\|T_x(\cdot, 0)\|^2 +
C\int_0^t\int_0^s\|\omega(\cdot, s_1)\|^2e^{-C'(s-s_1)/\lambda^2}ds_1ds\\
&&\le{C\lambda^2} \|T_x(\cdot, 0)\|^2+
{C\lambda^2}\int_0^t\|\omega(\cdot, s)\|^2ds.
\end{eqnarray*}
We use now the bound (\ref{bd-omegal2}) to get
\begin{eqnarray}\label{2.57}
&&\int_0^t\|T_x(\cdot, s)\|^2ds\le {C\lambda^2} \|T_x(\cdot, 0)\|^2+
\frac{C\lambda^2}{\sigma}\|v(\cdot, 0)\|^2+
{C\rho^2\lambda^6}\int_0^t\|T_x(s)\|_2^2ds.
\end{eqnarray}
Therefore there exists a constant $C_0>0$ so that if (\ref{R-bd}) holds, then
\[
\int_0^t\|T_x(\cdot, s)\|^2ds\le
{C\lambda^2} \|T_x(\cdot, 0)\|^2+
\frac{C\lambda^2}{\sigma}\|v(\cdot, 0)\|^2.
\]
Then (\ref{R-bd}) and (\ref{bd-omegal2}) imply that
\begin{equation}\label{2.7}
\int_0^t\|\omega(\cdot, s)\|^2ds\le
{C\rho^2\lambda^4} \left[{C\lambda^2}\|T_x(\cdot, 0)\|^2+
\frac{C\lambda^2}{\sigma}\|v(\cdot, 0)\|^2\right]+
\frac{C}{\sigma}\|v(\cdot, 0)\|^2
\end{equation}
$$
\le C\|T_x(\cdot, 0)\|^2+
\frac{C}{\sigma}\|v(\cdot, 0)\|^2.
$$
The bounds (\ref{2.6}) and (\ref{2.7}) together imply in an elementary
way that
\[
\|T_x(\cdot, t)\|\to 0~\hbox{as $t\to+\infty$.}
\]
We multiply the vorticity   equation (\ref{omegane}) by $\omega$
and integrate in space:
\begin{equation}\label{2.8}
\frac 12 \frac{d}{dt}\left(\|\omega(\cdot, t)\|^2\right)+\|\nabla\omega(\cdot, t)\|^2
\le \sigma\rho\|\omega(\cdot, t)\|\|T_x(\cdot, t)\|.
\end{equation}
Then we have
\begin{eqnarray*}
&&\|\omega(\cdot, t)\|^2\le\|\omega(\cdot, \tau)\|^2 +
2\sigma\rho\int_\tau^t\|\omega(\cdot, s)\|\|T_x(\cdot, s)\|ds\\
&&~~~~~~~~~~\le
\|\omega(\cdot, \tau)\|^2+2\sigma\rho\left(\int_\tau^\infty\|\omega(\cdot, s)\|^2ds\right)^{1/2}
\left(\int_\tau^\infty\|T_x(\cdot, s)\|^2ds\right)^{1/2}.
\end{eqnarray*}
The uniform bound (\ref{2.7}) implies that there exists a sequence of
times $\tau_k\to +\infty$ so that $\|\omega(\cdot, \tau_k)\|\to 0$. Then we
have for $t>\tau_k$:
\[
\|\omega(\cdot, t)\|^2\le\|\omega(\cdot, \tau_k)\|^2 +
2\sigma\rho\left(\int_{\tau_k}^\infty\|\omega(\cdot, s)\|^2ds\right)^{1/2}
\left(\int_{\tau_k}^\infty\|T_x(\cdot, s)\|^2ds\right)^{1/2}
\]
and hence $\|\omega(\cdot, t)\|\to 0$ as $t\to +\infty$ because of (\ref{2.57})
and (\ref{2.7}). In order to prove (\ref{kpp-upperbd}) we note that (\ref{2.8})
implies that
\[
\int_{t_1}^{t_2}\|\nabla\omega(\cdot, s)\|^2ds\le
\int_{t_1}^{t_2}\sigma\rho\|\omega(\cdot, s)\|\|T_x(\cdot, s)\|ds+\frac{1}{2}\|\omega(\cdot, t_1)\|^2
\]
so that (\ref{eq-thm3}) implies that for any $\eps>0$ we have for
$t_1$, $t_2>t_1+1$ sufficiently large we have
\[
\int_{t_1}^{t_2}\|\nabla\omega(\cdot, s)\|^2ds\le\eps(t_2-t_1).
\]
Therefore, (\ref{vineq}), (\ref{emb}) and the Cauchy-Schwartz inequality
imply that
\[
\bar V(t)\le 2+\frac{1}{t}\int_0^t\|w(s)\|_\infty ds+o(1)\le
2+\frac{C}{\sqrt{t}}\left(\int_0^t\|\nabla\omega(\cdot, s)\|^2ds\right)^{1/2}+o(1)
\le 2+\eps+o(1)
\]
and (\ref{kpp-upperbd}) follows.
This finishes the proof of Theorem \ref{thm3}.

\section{Linear instability}
\label{sec5}

In the previous section we established the stability of planar fronts
with respect to short wavelength perturbations. We analyze now the linear
instability of planar front with respect to long wavelength perturbations.

We perform a Galilean transformation
$z\mapsto z-v_0t$ in (\ref{ueq})-(\ref{teq}) following the flat front. We write
$T(x,z,t) = \tau(z-v_0t) + \theta (x, z-v_0t,t)$, and, with a slight abuse of
 notation, $v(x,z,t) = v(x, z-v_0t, t)$. We also linearize equations
 (\ref{ueq})-(\ref{teq}),
dropping the terms that are quadratic in $\theta$ and $v$. We
obtain the linearized system
\begin{eqnarray}
&&\pdr{\theta}{t} - v_0\partial_z\theta -\kappa \Delta \theta - \frac{v_0^2}
{4\kappa}f^{\prime}(\tau(z))\theta = - w\tau^{\prime}\la{thetaeq}\\
&&\pdr{u}{t} - v_0\partial_z u - \nu\Delta u + \nabla p = gA\theta e_z
\la{ulineq}
\end{eqnarray}
and for the vorticity we get the equation
\be
\pdr{\omega}{t} - v_0\partial_z\omega -\nu\Delta\omega = gA\partial_x\theta\la{omegaeq}.
\ee
Using the same reaction units (\ref{delta})-(\ref{thetaunits}) as before,
equations (\ref{thetaeq}), (\ref{omegaeq}) become
\be
\pdr{\theta}{t} -2\pdr{\theta}{z} - \Delta \theta - f^{\prime}(P(z))\theta =
-wP^{\prime}(z)\la{thetaqn}
\ee
and
\be
\pdr{\omega}{t} -2\pdr{\omega}{z} - \sigma \Delta \omega = \sigma\rho\pdr{\theta}{x}\la{omegaeqn}
\ee
The nonlinear equations, in the same units and frame of reference, are
\be
\pdr{\omega}{t} + v\cdot\nabla\omega - 2\pdr{\omega}{z} -\sigma\Delta\omega
= \sigma\rho\pdr{T}{x}\la{omegan}
\ee
and
\be
\pdr{T}{t} + v\cdot\nabla T - 2\pdr{T}{z} - \Delta T = f(T).\la{teqn}
\ee

We wish to show that flat profiles cannot remain linearly stable
with respect to sufficiently long wave infinitesimal perturbations.
We start for simplicity with the
case of the infinite Prandtl number. We retain (\ref{thetaqn})
but we use the infinite Prandtl number limit of (\ref{omegaeqn}),
\be
-\Delta \omega = \rho \pdr{\theta}{x},\label{infp}
\ee
which, together with (\ref{womega}) allows us to write the active scalar rule
\be
w = -\rho(\partial_x)^2(-\Delta)^{-2}\theta\la{ac}.
\ee
We express $\theta(x,z,t)$ in terms of its Fourier series:
\be
\theta (x,z,t) = \sum_{k\in \frac{2\pi}{\lambda}{\mathbf Z}}g_{k}(z,t)e^{ikx}.
\la{gk}
\ee
The linearized temperature equation (\ref{thetaqn}) transforms into
\be
\pdr{g_k}{t} - 2\pdr{g_k}{z} + \left (k^2 - \partial_{zz}\right )g_k
-f^{\prime}(P)g_k = \rho QKg_k\la{geq}
\ee
with $k = \pm\frac{2\pi}{\lambda}, \pm 2\frac{2\pi}{\lambda}, \dots$,
the operator $K$ defined by the Fourier transform of (\ref{ac})
\be
Kg =  k^2\left(k^2- \partial_{zz}\right)^{-2}g\la{K}
\ee
and
\be
Q(z) = - P^{\prime}(z)> 0.\la{Q}
\ee
We will choose the smallest positive wave number from now on:
\be
k = \frac{2\pi}{\lambda}\la{k}.
\ee
The operator $K$ defined by (\ref{K}) is given explicitly
by a convolution with a positive function
\be
(Kg) (z) =
\frac{1}{4k}\int_{-\infty}^{\infty}(1+k|z-\zeta |)e^{-k|z-\zeta|}g(\zeta)d\zeta.
\la{op}
\ee
The expression (\ref{op}) is obtained by an elementary calculation, starting from
\be
\left (k^2 -\partial_{zz}\right)^{-1}g (z) =
\frac{1}{2k}\int_{-\infty}^{\infty}e^{-k|z-\zeta |}g(\zeta)d\zeta \la{el}
\ee
and iterating.
It is well known that the profile $P$ is decreasing in the case of
KPP, bistable and ignition nonlinearities \cite{VVV},
so that the function  $Q$ is positive. Moreover
\be
Q(z) \ge ae^{-b|z|}\la{ca}
\ee
holds for all $z$, with $a>0$ and $b>0$ absolute numbers that depend only
on the nonlinearity $f(T)$. The positivity
of $Q$ and of the kernel of the operator $K$
in the right hand side of (\ref{geq}) imply that the
solution of the initial value problem (\ref{geq}) remains nonnegative if the
initial data is nonnegative. Let us consider
a function $\phi(z)$ which has the properties
\be
 e^{-k|z|}\le \phi (z)\le Ce^{-k|z|}\la{phione}
\ee with $C>1$ and \be |\phi^{\prime}(z)| \le Ck
e^{-k|z|},\,\,\,\, |\phi^{\prime\prime}(z)|\le Ck^2
e^{-k|z|}.\la{phitwo} \ee Such a function is obtained for instance
by gluing smoothly $3e^{1-z}$ on $z\ge 1$ to $3e^{1+z}$ on $z\le
-1$ by a function bounded below by $2$ on the interval $[-1,1]$.
Then one rescales $z\mapsto kz$.  We multiply the equation
(\ref{geq}) by $\phi(z)$  and integrate. Using the properties of
$\phi$ and the positivity of $g_k(z,t)$ we obtain \be
\frac{d}{dt}\int \phi(z)g_k(z,t)dz \ge - M\int\phi(z)g_k(z)dz +
\rho\int\phi(z)Q(z)(Kg_k)(z)dz\la{ine} \ee This follows from
integration by parts, and bounds (\ref{phitwo}) on the first two
derivatives of $\phi$ by constant multiples of $\phi$. The
constant $M\le 2C(1+k+k^2)$, so it is uniformly bounded for
bounded $k$, for instance for $0\le k\le 1$. Now we bound from
below the term \be \rho\int\phi(z)Q(z)(Kg_k)(z)dz \ge
\frac{\rho}{4k}\int\int ae^{-b|z|}\phi(z) e^{-k|z|-k|\zeta
|}g_k(\zeta, t)d\zeta dz\la{inter} \ee We used (\ref{ca}),
(\ref{op}) and the positivity of $g_k$. We neglected the
nonnegative term contributed by $k|z-\zeta |$ in (\ref{op}). Now
we use (\ref{phione}): \be \rho\int\phi(z)Q(z)(Kg_k)(z)dz \ge
\frac{\rho}{4k} \frac{a}{C}\left (\int
e^{-(b+2k)|z|}dz\right)\left (\int\phi(\zeta)g_k(\zeta,
t)d\zeta\right).\la{interm} \ee We obtain the ordinary
differential inequality \be \frac{d}{dt}\int \phi(z)g_k(z)dz \ge
\left (\frac{\rho}{4k}\frac{a}{2C(b+2k)} -M\right )
\int\phi(z)g_k(z)dz, \la{bineq} \ee and thus
$\|g_k\|_{L^1({\mathbb R})}$ grows exponentially in time.
Therefore we have the following theorem for the infinite Prandtl
number case:
\begin{thm}
Let $P(z-2t)$, $u =0$ be a planar,  $x$-independent traveling
front solution of the infinite Prandtl number Boussinesq system
\be \pdr{T}{t} + v\cdot\nabla T -\Delta T = f(T)\la{infprnoneq}
\ee with \be -\Delta v + \nabla p  = \rho {T}e_z,\quad \nabla\cdot
u = 0, \la{act} \ee with front boundary conditions for $T$ at
$z=\pm\infty$, vanishing velocity at $z= \pm \infty$ and periodic
boundary conditions in $x$ of period $\lambda$. There exists a
positive constant $\beta>0$ such that, if \be \rho \lambda >
\beta, \la{cond} \ee then the solution $P$ is linearly unstable.
This means that there exist infinitesimal perturbations which grow
exponentially, when viewed in a Galilean frame of reference moving
with the traveling front. Their exponential growth rate is
proportional to $\rho\lambda$.
\end{thm}
A similar instability analysis may be applied to a convective system with the
infinite Prandtl number equation (\ref{infp}) replaced by the Darcy
law that was studied numerically in \cite{deWit}.

We return to the system (\ref{thetaqn}, \ref{omegaeqn}) at a
finite Prandtl number. We use the Fourier expansion \be w(x, z, t)
= \sum_{n\in{\mathbf{Z}}}w_k(z,t)e^{{2\pi inx}/{\lambda}} \la{voo}
\ee and using (\ref{womega}) we have, with $k = {2\pi
n}/{\lambda}$, \be w_k(z,t) = -\frac{i}{2}\int
e^{-k|z-z^{\prime}|}\omega_k(z^{\prime}, t)dz^{\prime}\la{pizza}
\ee with the obvious notation for the Fourier coefficients of
$\omega$. An elementary calculation involving solving the linear
equation (\ref{omegaeqn}) starting with the zero initial data
$\omega(x,z,0)=0$, and using (\ref{womega}), gives \be w_k(z,t) =
\frac{\sigma\rho k}{2} \int_0^t\int g_k(z_2,s)I_k(z-z_2 +
2(t-s),t-s)dz_2ds\la{wk} \ee where $g_k(z,t)$ are the Fourier
coefficients of  $\theta$ as in (\ref{gk}) and \be I_k(\zeta,\tau)
= e^{-\sigma k^2\tau}\int e^{-k|\zeta  + u\sqrt{2\sigma \tau}|}
e^{-\frac{|u|^2}{2}}\frac{du}{\sqrt{2\pi}}.\la{I} \ee We now bound
$I_k$ at this point as follows:
\be I_k(\zeta,\tau)\ge e^{-\sigma k^2\tau}\int e^{-k|\zeta|
-k|u|\sqrt{2\sigma \tau}}
e^{-\frac{|u|^2}{2}}\frac{du}{\sqrt{2\pi}}\ge
2e^{-k|\zeta|}\Erf(k\sqrt{2\sigma\tau})\la{erf} \ee with \be
\Erf(a) = \int_a^{\infty}e^{-\frac{u^2}{2}}\frac{du}{\sqrt{2\pi}}.
\la{erfa} \ee For $a\ge 1$ we  will use just a piece of this
integral, \be \Erf(a) \ge
\int_a^{2a}e^{-\frac{u^2}{2}}\frac{du}{\sqrt{2\pi}}\ge
\frac{1}{\sqrt{2\pi}}e^{-a^2}\la{erfi} \ee and for $a\le 1$ we
write \be \Erf(a)\ge \Erf(1).\la{obv} \ee This allows us to
deduce, for later use \be I_k(\zeta,\tau)\ge
2(1-\frac{1}{\sqrt{2\pi}}) e^{-k|\zeta |} e^{-2\sigma
k^2\tau}.\la{ii} \ee

The evolution equation for $g_k$ follows from (\ref{thetaqn}),
\be
\pdr{g_k}{t} - 2\pdr{g_k}{z} + (k^2 - \partial_{zz})g_k
-f^{\prime}(P)g_k = \sigma\rho Qw_k\la{gkeqs}
\ee
with $w_k$ computed using (\ref{wk}) and with $Q=-P^{\prime}$ as before.
Because of the explicit expression (\ref{wk}) we see that if the
initial data $g_k(z,0)$ is real and positive, then $g_k(z,t)$ remains real
and positive. We take such initial data. Then, using (\ref{ii}) and the
triangle inequality we get,
\be
w_k(z,t) \ge (1-\frac{1}{\sqrt{2\pi}})\rho\sigma k e^{-k|z|}\int_0^te^{-2(t-s)k(\sigma k + 2)}\left
[\int g_k(z_2,s)e^{-k|z_2|}dz_2\right]ds.\la{wkneq}
\ee
As in the case of the infinite Prandtl number, we take a function $\phi(z)$
with properties (\ref{phione}), (\ref{phitwo}), multiply (\ref{gkeqs})
by $\phi$ and  integrate.
Let us denote
\be
Y(t) = \int \phi(z)g_k(z,t)dz.\la{yt}
\ee
Integrating by parts we get
\be
\frac{dY}{dt} + C_3(1+k +k^2)Y \ge C_4\sigma\rho k\int_0^te^{-2(t-s)k(\sigma k + 2)}Y(s)ds.\la{int}
\ee
with $C_3$ and $C_4$ positive numbers that depend on the properties of
$\phi$. Consider also
\be
Z(t) = e^{2k(\sigma k +2)t}Y(t).\la{zt}
\ee
Multiplying by $e^{2k(\sigma k+2)t}$, we get from (\ref{int})
\be
\frac{dZ}{dt} + \beta Z \ge \alpha \int_0^tZ(s)ds\la{zeq}
\ee
with
\be
\beta = C_3(1+k+k^2) -2k(\sigma k +2)\la{b}
\ee
and
\be
\alpha = C_4\sigma\rho k.
\la{alp}
\ee
It is clear that solutions of the   differential
inequality (\ref{zeq}) are larger than solutions of the
ODE
\be
\frac{dy}{dt} +\beta y = \alpha\int_0^ty(s)ds
\la{yone}
\ee
with the same initial data. This ODE is solved differentiating one
more time,
\be
\frac{d^2 y}{dt^2} + \beta\frac{dy}{dt} -\alpha y = 0\la{ytwo}
\ee
and seeking solutions of the form $e^{\mu t}$.
Equation (\ref{ytwo}) always has at least one exponentially growing solution
because $\alpha>0$ (irrespective of the sign of $\beta$).
The general solution of (\ref{ytwo}) is
\be
y(t) = y_1e^{t\mu_{+}} + y_2e^{t\mu_{-}}\la{gy}
\ee
with
\be
\mu_{\pm} = \frac{1}{2}\left ( -\beta \pm {\sqrt{\beta^2 +4\alpha}}\right ).\la{mus}
\ee
The solutions of (\ref{yone}) correspond to the linear subspace for which
$\frac{dy}{dt}(0) + \beta y(0) = 0$. Choosing without loss of generality the
coefficient of the growing exponential $y_1 = 1$, we deduce the relation
\be
\mu_{+} + \beta + y_2(\mu_{-} + \beta) = 0\la{rel}
\ee
In order to have exponentially growing positive solutions we need
the initial datum for (\ref{yone}) to be positive. This initial
datum is $1 + y_2$ and is computed using (\ref{mus}) and (\ref{rel})
\be
1+y_2 = \frac{2\sqrt{\beta^2 + 4\alpha}}{\sqrt{\beta^2 + 4\alpha}-\beta}
\la{id}
\ee
so it is always positive. Starting with this initial condition (or any
positive multiple thereof) we get exponential growth for
$Z(t)$. This will imply exponential growth for $Y(t)$ if
\be
\mu_{+} > 2k(\sigma k + 2).\la{exps}
\ee
This turns out to be the condition
\be
C_4\sigma \rho > 2C_3(1+k+k^2)(\sigma k +2)\la{endp}
\ee
or, putting $C=\frac{2C_3}{C_4}$,
\be
\rho > C\left (\frac{2}{\sigma} + k\right )\left (1 + k + k^2\right )\la{endpr}
\ee
\begin{thm} Planar reactive Boussinesq fronts (\ref{omegan}, \ref{teqn})
are linearly unstable to large wavelength perturbations whenever the local
Rayleigh number $\rho$  based
on the laminar front thickness is large compared to the inverse of the
Prandtl number,
$$
\rho > \frac{2C}{\sigma}.
$$
Perturbations with wave numbers $k$ satisfying (\ref{endpr}) grow
exponentially in a frame of reference moving with the planar
front. The growth rate is proportional to
$\sqrt{\sigma\rho k}$.
\end{thm}
\noindent{\bf{Remark.}} The exponential
growth rate proportional to the square root of the wave number is a signature
of the Rayleigh-Taylor instability, operating here only at large scales. When
the wave length of the initial perturbation is decreased to a length
comparable to the thickness of the planar front, the perturbation decays in
time.
\section{Conclusions}
We proved that the reactive Boussinesq system in a strip has front
solutions with bounded speeds, if the nonlinearity is of concave
KPP type. The bounds grow with the aspect ratio, but nevertheless,
front acceleration does not occur in this system. The concavity of
the nonlinearity was used in the proof. We do not know at present
whether the divergence of the bound with aspect ratio is sharp.
For small aspect ratios and for small Rayleigh numbers, the only
traveling modes are planar, and all front-like solutions become
planar. This result is expected to be valid in more generality
than proved here. For large enough Rayleigh numbers, if the aspect
ratio is large, then the planar fronts lose
 stability to long wave perturbations. The instability is of Rayleigh-Taylor
type. This again is expected to hold in greater generality than
proved here. The results proved here agree qualitatively with the
recent numerical study \cite{VR}.

\vspace{.5cm}

\noindent{\bf Acknowledgments} The authors thank N. Vladimirova
and R. Rosner for extensive scientific interactions. PC was
supported partially by NSF DMS-0202531. AK was supported partially
by NSF grant DMS-0102554 and by an Alfred P. Sloan Research
Fellowship. LR was supported partially by NSF grant DMS-0203537
and by an Alfred P. Sloan Research Fellowship. This research is
supported in part by the ASCI Flash center at the University of
Chicago under DOE contract B341495.


\begin{thebibliography}{99}

\bibitem{B1} H.~Berestycki, The influence of advection on 
the propagation of fronts in
reaction-diffusion equations, Proceedings NATO ASI Conf. Cargese,
H. Berestycki and Y. Pomeau eds. Kluwer (to appear).

\bibitem{BH} H.~Berestycki and F.~Hamel, Front propagation in
periodic excitable media,  Comm. Pure Appl. Math. {\bf 55}, 2002, 949--1032.

\bibitem{BLL} H. Berestycki, B. Larrouturou and P. L. Lions,
 Multi-dimensional traveling wave solutions of a flame
propagation model, Arch. Rational Mech. Anal., {\bf 111},
1990, 33-49.

\bibitem{BN} H. Berestycki and L. Nirenberg,  Traveling fronts in
cylinders,  Annales de l'IHP, Analyse non lin\'eare, {\bf 9},
1992, 497-572.

\bibitem{CKOR} P. Constantin, A. Kiselev, A. Oberman, L. Ryzhik, Bulk
burning rate in passive-reactive diffusion, Arch. Rat. Mech. Anal. {\bf 154},
2000, 53-91.

\bibitem{CKR} P. Constantin, A. Kiselev, L. Ryzhik, Quenching of flames by
fluid advection, Comm. Pure Appl. Math {\bf 54}, 2001, 1320-1342.

\bibitem{chandra} S. Chandrasekhar, {\it 
Hydrodynamic and hydromagnetic stability},
Clarendon Press 1961.

\bibitem{DM1} J. Daou and M. Matalon, Flame propagation in Poiseuille 
flow under adiabatic conditions, Combustion and Flame, {\bf  124}, 2001,
337-349.

\bibitem{DM2} J. Daou and M. Matalon, Influence of conductive 
heat-losses on the propagation of premixed flames in channels,
Combustion and Flame {\bf 128}, 2001, 321-339.

\bibitem{deWit} A. de Wit, Fingering of chemical fronts in porous media,
Phys. Rev. Let., {\bf 87}, 054502.

\bibitem{F1} M. Freidlin and J.~G\"artner,  On the propagation
of concentration waves in periodic and random media, Soviet
Math. Dokl., {\bf 20}, 1979, 1282-1286.

\bibitem{F2} M.~Freidlin,  Geometric optics approach to
reaction-diffusion equations,  SIAM J. Appl. Math., {\bf 46},
1986, 222-232.

\bibitem{HPS} S.~Heinze, G.~Papanicolau and A.~Stevens, 
Variational principles for propagation speeds in inhomogeneous
media,  SIAM J. Appl. Math. {\bf 62}, 2001, 129-148.

\bibitem{KS} L.~Kagan and G.~Sivashinsky, Flame propagation and extinction 
in large-scale vortical flows, Combust. Flame {\bf 120}, 2000, 222-232.

\bibitem{KRS}  L.~Kagan, P.D.~Ronney and G.~Sivashinsky, 
Activation energy effect on flame propagation in large-scale vortical
flows, \rm Combust.  Theory Modelling {\bf 6}, 2002, 479-485.

\bibitem{KR} A.~Kiselev and L.~Ryzhik, {Enhancement of the travelling front
speeds in reaction-diffusion equations with advection}, Ann. Inst.
H. Poincar\'e Anal. Non Lin\'eaire {\bf 18}, 2001, 309-358.

\bibitem{MS}
A.~Majda and P.~Souganidis, Large scale front dynamics for turbulent
reaction-diffusion equations with separated velocity scales,
Nonlinearity, {\bf 7}, 1994, 1-30.

\bibitem{MX} S. Malham and J. Xin, Global solutions to a reactive Boussinesq
system with front data on an infinite domain, Comm. Math. Phys., {\bf 193}, 
1998, 287-316.

\bibitem{Peters} N.~Peters, \it Turbulent Combustion, \rm
Cambridge University Press, Cambridge, UK 2000.

\bibitem{TV1} R.~Texier-Picard and V.~Volpert, Probl\`emes de
r\'eaction-diffusion-convection dans des cylindres non born\'es,
C. R. Acad. Sci. Paris Sér. I Math. {\bf 333}, 2001, 1077-1082

\bibitem{TV2} R.~Texier-Picard and V.~Volpert,
Reaction-diffusion-convection problems in unbounded cylinders, \rm
Preprint 2001.

\bibitem{VR} N. Vladimirova, R. Rosner, Model flames in the Boussinesq limit:
the effects of feedback, Preprint, 2002.

\bibitem{VCKRR} N.~Vladimirova, P.~Constantin, A.~Kiselev,
O.~Ruchayskiy and L.~Ryzhik, in preparation.


\bibitem{VVV} A.~Volpert, V.~Volpert and V.~Volpert, Travelling Wave Solutions
of Parabolic Systems, Translations of mathematical Monographs,
{\bf 140}, Amer. Math. Soc., Providence, Rhode Island 1994.

\bibitem{X1} J.~Xin, Existence of planar flame fronts in
convective-diffusive periodic media, Arch. Rat. Mech. Anal., {\bf
121}, 1992, 205-233.

\bibitem{X2} J.~Xin, Existence and nonexistence of travelling waves
  and reaction-diffusion front propagation in periodic media, 
  Jour.  Stat. Phys., {\bf 73}, 1993, 893-926.

\bibitem{X3} J.~Xin, Analysis and modelling of front propagation in
heterogeneous media, SIAM Rev., {\bf 42}, 2000, 161-230.

\bibitem{ZBLM} Ya.B.~Zeldovich, G.I.~Barenblatt, 
V.B.~Librovich and G.M.~Makhviladze,
\it The Mathematical Theory of Combustion and Explosions, \rm
Translated from the Russian by Donald H. McNeill. Consultants
Bureau [Plenum], New York, 1985.

\end{thebibliography}
\end{document}